\documentclass[prd,aps,preprint,tightenlines,showpacs,nofootinbib,superscriptaddress]{revtex4-1}
\usepackage{mathrsfs}
\usepackage{amsfonts}
\usepackage{amsmath}
\usepackage{amssymb}
\usepackage{array}
\usepackage{verbatim}
\usepackage{bm}
\usepackage{epsfig}
\usepackage{graphicx,color}
\usepackage{relsize}
\usepackage{lineno}
\usepackage{float}
\usepackage{multirow}
\RequirePackage{xspace}
\begin{document}

\title{Exclusive $J/\psi$ production in proton-proton collisions adopting GPD approach}
\author{Ya-Ping Xie}
\email{xieyaping@impcas.ac.cn}
\affiliation{Institute of Modern Physics, Chinese Academy of Sciences, Lanzhou 730000, China}
\affiliation{University of Chinese Academy of Sciences, Beijing 100049, China}
\affiliation{State Key Laboratory of Heavy Ion Science and Technology, Institute of Modern Physics,
	Chinese Academy of Sciences, Lanzhou 730000, China}
\author{ S.V.Goloskokov}
\affiliation{Laboratory of Theoretical Physics, Joint Institute for Nuclear Research, Dubna 141980, Moscow region, Russia}
\affiliation{Institute of Modern Physics, Chinese Academy of Sciences, Lanzhou 730000, China}
\begin{abstract}
    Exclusive $J/\psi$ production is investigated in proton-proton collisions employing GPD approach with GK model. Three
    sets gluon density are used to calculate exclusive $J/\psi$ production. The survival factors and equivalent photon approximation 
    are applied to predict the exclusive $J/\psi$ photoproduction in proton-proton collisions. The GPD method prediction gives a good
     agreement with the experimental data at LHCb.
     The exclusive $J/\psi$ production is given in proton-proton collisions at RHIC and NICA. 
     These predictions of GPD approach can be employed to estimate the  exclusive $J/\psi$ cross sections
     in future proton-proton collisions.
\end{abstract}
\maketitle

\section{Introduction}\label{intro}
In hadron-hadron collisions, the exclusive heavy vector meson production can be adopted to study the hadron structure in 
high energy limit \cite{upc, upc02}. In hadron-hadron collisions, the photon-hadron scattering is important to produce exclusive 
vector meson production. 
The photon flux in hadron-hadron collisions is applied in equivalent photon approximation method\cite{twophotons}.
The vector mesons production at heavy-ion collisions have been measured in several collaborations.
 ALICE have measured exclusive $J/\psi$ rapidity
distributions at Pb-Pb and p-Pb ultraperipheral collisions\cite{ALICE:2013wjo} and LHCb have performed measurements of exclusive $J/\psi$ at
 proton-proton collisions\cite{LHCb:2013nqs,LHCb:2014acg, LHCb:2016oce}. PHENIX measured exclusive $J/\psi$ meson production at Au-Au
  ultraperipheral collisions\cite{PHENIX:2009xtn}. These measurements have been employed to investigate the nucleon structure property\cite{LHCUPC}.

Generalized parton distributions (GPDs) have provided a rigorous formulation of the spatial
distributions of quarks and gluons in the nucleon\cite{GPD-review01,GPD-review02}.
The GPDs enable the construction of 3D tomographic images of the distributions
of quarks and gluons\cite{Muller:1994ses,Ji:1996ek, Radyushkin:1997ki}.
Several physical processes can be adopted to study GPDs, for example, Deep Virtual Compton Scattering (DVCS)\cite{dvcs}, Time-like Compton
Scattering (TCS)\cite{TCS, Xie:2022vvl}, Deep Vector Meson Production (DVMP)\cite{dvmp1, Goloskokov:2022mdn}. In DVMP process, the
 light flavor meson production can be used to study quark and gluon GPDs.
The heavy meson production can be employed to study gluon GPDs in proton. Neutrino production can be also used to
study GPD in hadron\cite{Pire:2017tvv}.

Dipole model is important to study the exclusive vector meson production
in heavy-ions collisions and electron-proton scattering\cite{PRD68, PRD74, PRD78}. In dipole model, the photon can be viewed
as quark and anti-quark pair(dipole). In high energy, the virtual  photon initiate dipole state that interacts with gluons from the hadrons and produces
a vector meson in a final state. Dipole model have been adopted to predict vector meson production in hadrons
ultraperipheral collisions\cite{Lappi:2013am,Goncalves:2014swa, Xie:2016ino,Xie:2018tng}.
Even if the dipole model can describe the experimental data, there are some issues in the dipole model. Firstly, the dipole amplitude
is the fitted at small-x region ($x_B< 0.01$). Secondly, the skewness of the gluon can not be described in dipole model.  In our previous work, we have
 studied $J/\psi$ cross section employing GPDs approach in electron-proton scattering\cite{Goloskokov:2024egn}.
The predictions of our work can describe the experimental data well. Consequently, we can adopting the GPD method to study the  exclusive $J/\psi$ production  in proton-proton collisions.

In process of exclusive vector meson production in proton-proton collision at LHCb, the survival factor is introduced in 
theoretical investigation \cite{Khoze:2013jsa, Jones:2016icr}.
The survival factor is associated with soft rescatterings in proton-proton collisions that fills the rapidity gap. At LHCb, 
the final states of  proton can not able to be measured and the diffractive events are selected by measuring the rapidity gap. 
However, as the pile-up is high, the probability of soft rescatterings and the 
overlap of inelastic processes that generates additional particles in the rapidity region covered by the LHCb is also high.
 As a consequence, these events suppress the number of exclusive vector meson production  at the LHCb. 
In order to compare our predictions with the  LHCb data, we need to take into account this suppression effect.

 In hadron-hadron collisions, the two hadrons can be viewed as the source of the final states. Therefore, the final particles can interference in
 amplitudes. The differential cross sections is affected by the amplitude interference contributions\cite{Xing:2020hwh,Zha:2020cst}. 
 However, the interference contributions are not important
  in the total cross section of vector mesons, where they can be neglected in the calculations \cite{yajin-zhou}.
   In this work, we don't consider the interference contributions in rapidity distributions of vector mesons in proton-proton ultraperipheral collisions.

Nuclotron-based Ion Collider fAcility (NICA) is a new accelerator complex designed at the Joint Institute for Nuclear Research (Dubna, Russia) to study
properties of baryonic matter\cite{Sorin:2011zz}. The energy of NICA is low while Brojken x is large.
 Thus, the vector meson production
 can be investigated the nucleon property in large-x limit. Relativistic Heavy Ion Collider (RHIC) is a platform to study nuclear structure
 and nuclear matter physics. STAR and PHOENIX are two collaboration groups on RHIC \cite{STAR:2004jhn,PHENIX:2004vcz}.

 This paper is organized as follows. In Section II, the theoretical frame of exclusive vector meson production in proton-proton collisions are presented.
 In Section III, the numerical results are shown in figures. The predictions of exclusive $J/\psi$ at proton-proton collisions at LHCb are compared
 with experimental data. The prediction of exclusive $J/\psi$ at proton-proton collisions at RHIC and NICA  are also exhibited. Some conclusions and
 discussions are given in Section IV.
 \section{Vector meson cross section in proton-proton collisions}
 In hadron-hadron scattering process, the hadron can accur strong electro-magnetic (EM) field at high energy limit. The strong EM field can
 be viewed as quasi-real photon. When the impact parameter between two nucleus is large, the strong interaction in heavy ions can be neglected. The
 electro-magnetic interaction is dominant in this situation. In proton-proton collisions, the photon can interact with proton and produce vector
 mesons. The two beam can accure two photon flux which scatters off the proton beam and produces mesons.
 At the proton-proton collisions, the exclusive vector meson rapidity distributions can be written as follows\cite{Jones:2013pga,Jones:2016icr}
 \begin{eqnarray}
 \frac{d\sigma ^{th}(pp)}{dy} = S^2(W_{+})\Big(k_+\frac{dn}{dk_+}\Big)\sigma^{th}_+(\gamma p) +S^2(W_{-})\Big(k_-\frac{dn}{dk_-}\Big)\sigma^{th}_-(\gamma p).
 \end{eqnarray}
 $k$ is the photon momentum, $k_{\pm} =x_\gamma ^\pm \sqrt{s_{NN}}/2\approx (M_V/2)e^{\pm |\mathrm{y}|}$. $\mathrm{y}$ is the rapidity of vector meson and $M_V$ is the mass
 of the vector meson, $s_{NN}$ is the collisions energy and $W^2_{\pm}=M_V\sqrt{s_{NN}}e^{\pm |\mathrm{y}|}$.
 $S^2(W_\pm)$ is the survival factors which have been studied in Ref\cite{Khoze:2013jsa} and they can be found in Ref\cite{Jones:2016icr}.

In proton-proton collision, there are two amplitude
of photon-proton interaction. The two amplitudes can interference with each other when we consider the differential cross sections
distributions. However, the interference contributions happen at small $p_t$ region, when we consider the total cross sections\cite{Xing:2020hwh}, 
the interference cross section is small compare to total cross sections. We can neglect the interference
contributions when we consider the total cross sections. 

  In this work, we use the survival factors in LHCb proton-proton
  collisions since the collision energy is high. In the NICA and RHIC proton-proton collisions, we suppose that the survival
 factors equal to unity that is in accordance with Ref\cite{Jones:2016icr} calculations. 
 
 $dn/dk$ is the photon flux of the proton.
 The photon flux of the proton is given as\cite{upc02}
\begin{eqnarray}
\frac{dn}{dk}(k)=\frac{\alpha_{em}}{2\pi k}\Big[1+\Big(1-\frac{2k}{\sqrt{s_{NN}}}\Big)^2\Big]\Big(
\ln \Omega-\frac{11}{6}+\frac{3}{\Omega}-\frac{3}{2\Omega^2}+\frac{1}{3\Omega^3}\Big),
\end{eqnarray}
where $\Omega=1+0.71/Q^2_{min}$, with $Q^2_{min}=k^2/\gamma^2_L$, $\gamma_L$ is the Lorentz boost factor with $\gamma_L=\sqrt{s_{NN}}/2m_p$.
$\sigma^{th}_\pm(\gamma p)$ is the photon-proton scattering cross sections which can calculated via a couple of method, for instance, dipole model,
kt factorization and GPD approach. In previous paper\cite{Goloskokov:2024egn}, we adopt GPD approach to study the $J/\psi$ production in
electron-proton scattering.
The photon-proton scattering cross section is obtained as
\begin{eqnarray}
\frac{d\sigma}{dt}=\frac{d\sigma_T}{dt} \big[1+\frac{Q^2}{m_V^2}\big],\;\;\;\; \frac{d\sigma_T}{dt}=\frac{1}{16\pi W^2(W^2+Q^2) }\big[|\mathcal{M}_{++,++}|^2+|\mathcal{M}_{+-,++}|^2\big].
\end{eqnarray}
The total cross section can be obtained by integrating $|t|$ as
\begin{equation}
\sigma^{th}=\int_{t_{min}}^{t_{max}}\frac{d\sigma}{dt}\simeq \frac{1}{2B_V}\frac{d\sigma}{dt}(t=0).
\end{equation}
The slope factor $B_V$ is given as \cite{Goloskokov:2024egn}
\begin{equation}
B_V= b_0+b_1 \ln[m_V^2/(m_V^2+Q^2)]+\alpha' \ln[(W^2+Q^2)/(m_V^2+Q^2)].
\end{equation}
The amplitudes $\mathcal{M}$ in Eq.~(3) can be represented as a convolution of hard amplitude $\mathcal{H}^{V}$ that calculated perturbatively and
soft part gluon GPDs.
 The helicity conservation amplitude
can be written as \cite{gk06}
\begin{eqnarray}
\mathcal{M}_{\mu^\prime+,\mu+}
&=& \frac{e}{2}C_V\int_0^1\frac{dx}{(x+\xi)(x-\xi+i\epsilon)}\mathcal{H}^{V+}_{\mu^\prime,\mu}\,H_g(x, \xi, t).
\end{eqnarray}
On the other hand, the helicity flip amplitude is given as
\begin{eqnarray}
\mathcal{M}_{\mu^\prime-,\mu+}
&=&-\frac{e}{2}C_V\frac{\sqrt{-t}}{2m}\int_0^1\frac{dx}{(x+\xi)(x-\xi+i\epsilon)}\mathcal{H}^{V+}_{\mu^\prime,\mu}\,E_g(x, \xi, t).
\end{eqnarray}
We omit here the contributions from polarized gluon GPDs that is unimportant here, $\mathcal{H}^{V+}_{\mu^\prime,\mu}$ contains the sum of vector meson production amplitudes with different gluon helicities.
The detail information of the amplitudes calculations can be found in our previous paper \cite{Goloskokov:2024egn}. The GK model will be adopted in this
 work\cite{gk07,gk08,gk09}. GPDs in this model are connected with PDFs with the help of the double distribution representation \cite{mus99} that reconstructed the skewness $\xi$ dependencies of GPDs. Using Eq.~(6) and Eq.~(7), we calculate real and imaginary parts of amplitudes $\mathcal{M}$ that contribute to the differential cross section.
Three gluon density sets \cite{Alekhin:2018pai,Hou:2019qau,H1:2015ubc} are employed in constructing the GPD.

It can be seen that we only consider the Leading Order (LO) exclusive vector meson contribution in photon-proton scattering process. 
Next Leading Order (NLO) cross section in GPD method were studied in e.g. Ref\cite{Flett:2021ghh}. It was found that LO results exceed
NLO contribution by factor $2\sim3$. We observed the same effect in our model by taken into account the transverse part momenta of 
quark momenta  (see corresponding discussion for
light vector meson production in Ref\cite{gk06} and for $J/\psi$ production in Ref\cite{Goloskokov:2024egn}).
 \section{Numerical results}
The exclusive $J/\psi$ production were measured at LHCb in proton-proton collisions\cite{LHCb:2014acg,LHCb:2016oce}.
 Some theoretical predictions can be found at Refs\cite{Jones:2013pga,Jones:2016icr,Xie:2018tng}.
 In this paper, we adopt GPD calculations to give predictions for $J/\psi$ production in this reaction and
 compare them with the experimental data. The survival
 factors and equivalent photon approximation, as discussed in section II, are employed in our calculations. The values of the survival factors can be found at
  Ref.\cite{Jones:2016icr}. The $J/\psi$ photoproduction is calculated for three sets of GPDs, based on the NLO gluon densities
   \cite{Alekhin:2018pai,Hou:2019qau,H1:2015ubc}.

 First of all, the production of $J/\psi$ as a function of the vector meson rapidity in proton-proton collisions at LHCb
 are shown in Fig.~1 for the CT18 set of the gluon density \cite{Hou:2019qau}.
   It can be seen that our central results at $\sqrt{s_{NN}}$ = 7 TeV (left graph) and at $\sqrt{s_{NN}}$ = 13 TeV (right graph)agree with the experimental data.
 The cross section uncertainties of CT18 is large since the gluon density uncertainties of CT18 is large at small-x range.
\begin{figure}[!h]
\includegraphics[width=6.9cm]{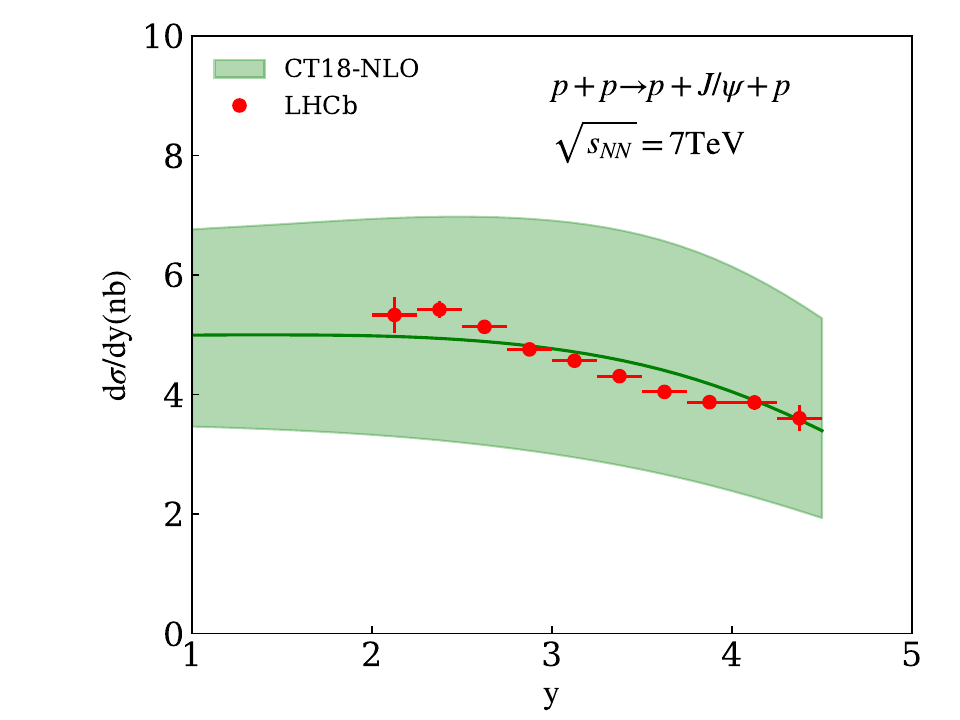}
\includegraphics[width=6.9cm]{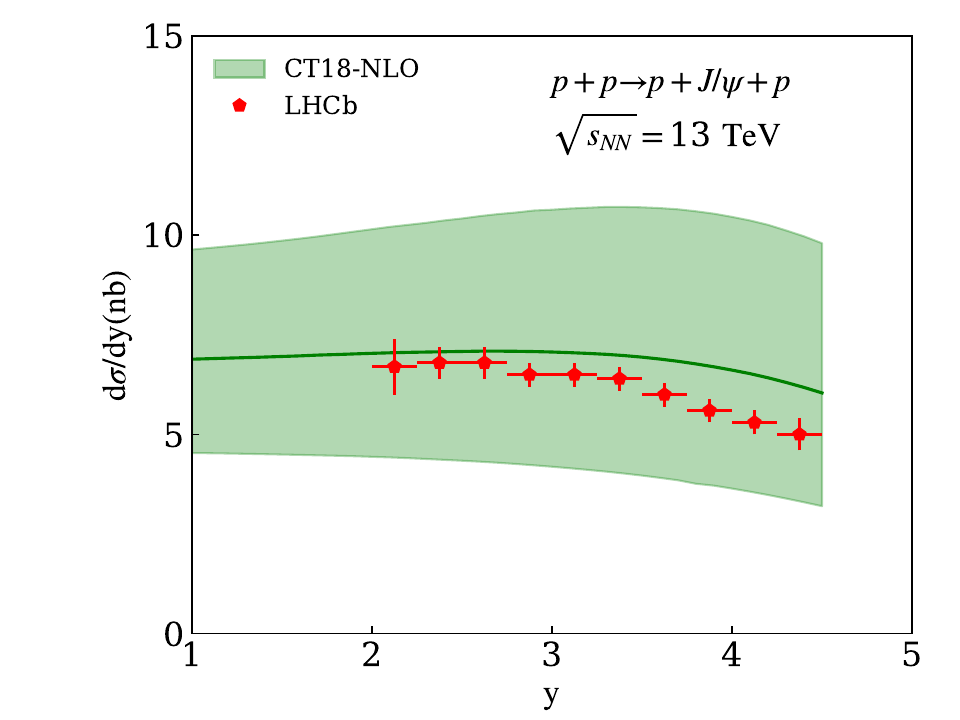}
\noindent\caption{Exclusive production of $J/\psi$ as a function of rapidity in proton-proton collisions and comparison with experimental data of LHCb\cite{LHCb:2014acg, LHCb:2016oce}. The gluon density is CT18 set\cite{Hou:2019qau}.}
\end{figure}

  The prediction of  exclusive $J/\psi$ cross section in proton-proton 
 collisions as a function rapidity for GPDs based on the ABMP16 gluon density \cite{Alekhin:2018pai} are exhibited in Fig.~2. It can seen that the
  ABMP16 prediction overestimate the experimental
 data. The uncertainties in this case are smaller with respect to shown in Fig.~1 because  the gluon density uncertainties of ABMP16  are smaller than CT18
  uncertainties at small-x limit.
\begin{figure}[h!]
	\includegraphics[width=6.9cm]{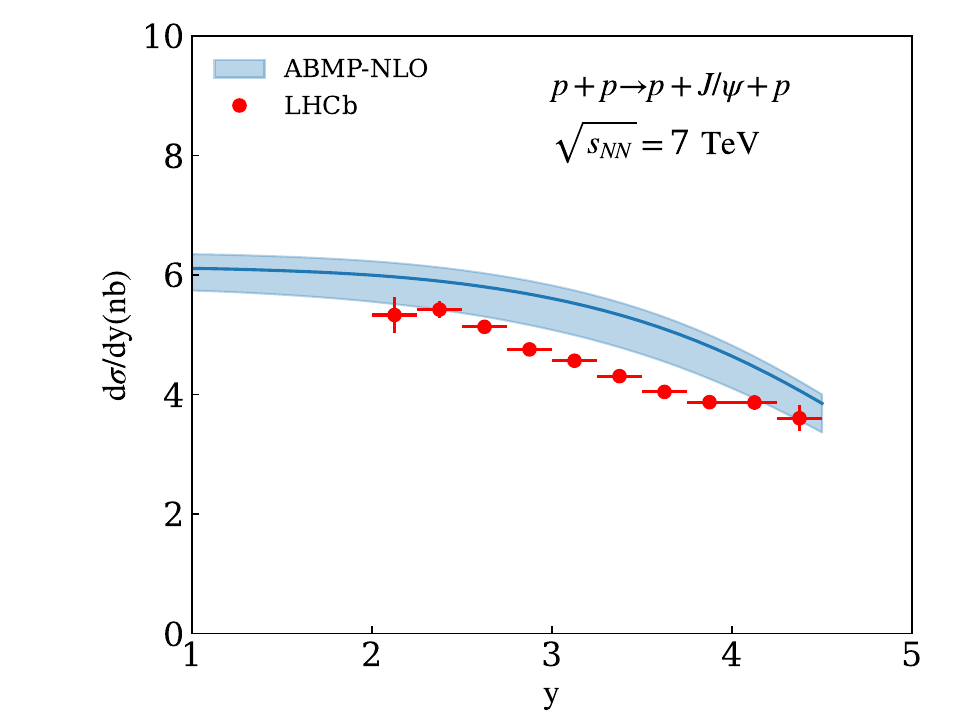}
	\includegraphics[width=6.9cm]{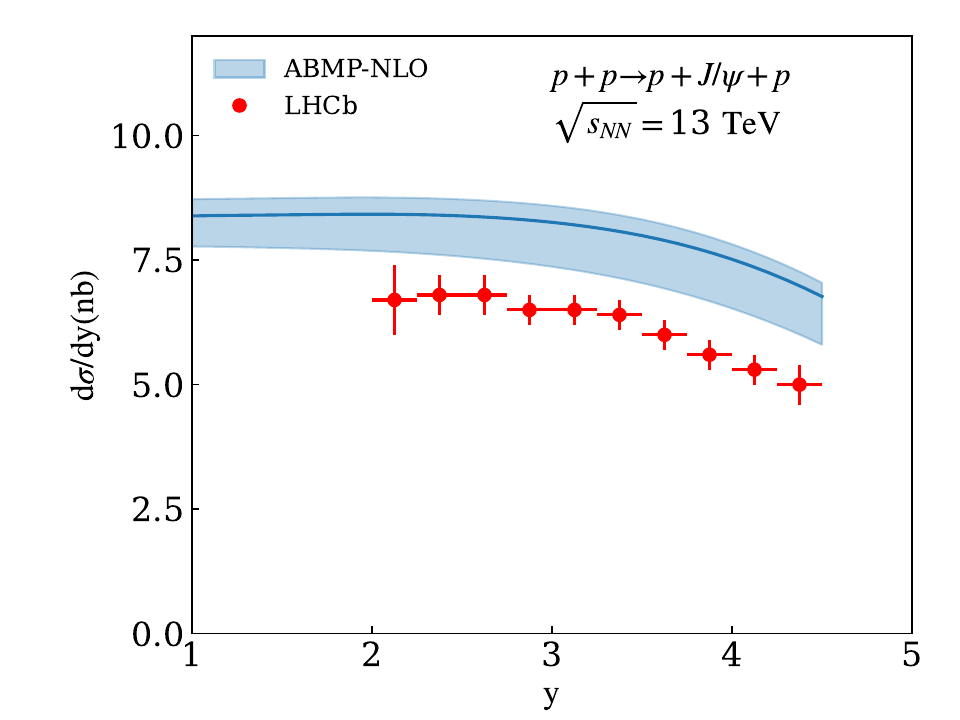}
	\noindent\caption{Exclusive production of $J/\psi$ as a function of rapidity in proton-proton collisions comparison with experimental data of LHCb\cite{LHCb:2014acg, LHCb:2016oce}. The gluon density is ABMP16 set\cite{Alekhin:2018pai}.}
\end{figure}

Our predictions of $J/\psi$ production for the  HERA15 gluon density \cite{H1:2015ubc} are
presented at Fig.~3. It can be seen that in this case the results describe the LHCb data well.
The uncertainties of HERA15 gluon density is smaller than
for CT18 that provide quite small uncertainties in the cross sections.
\begin{figure}[h!]
	\includegraphics[width=6.9cm]{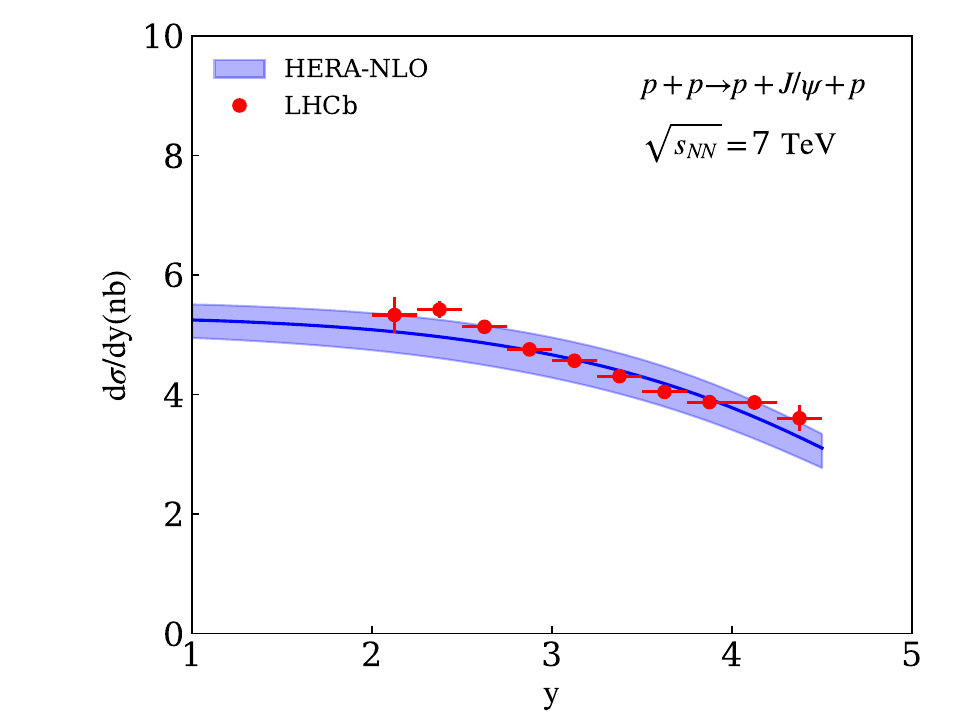}
	\includegraphics[width=6.9cm]{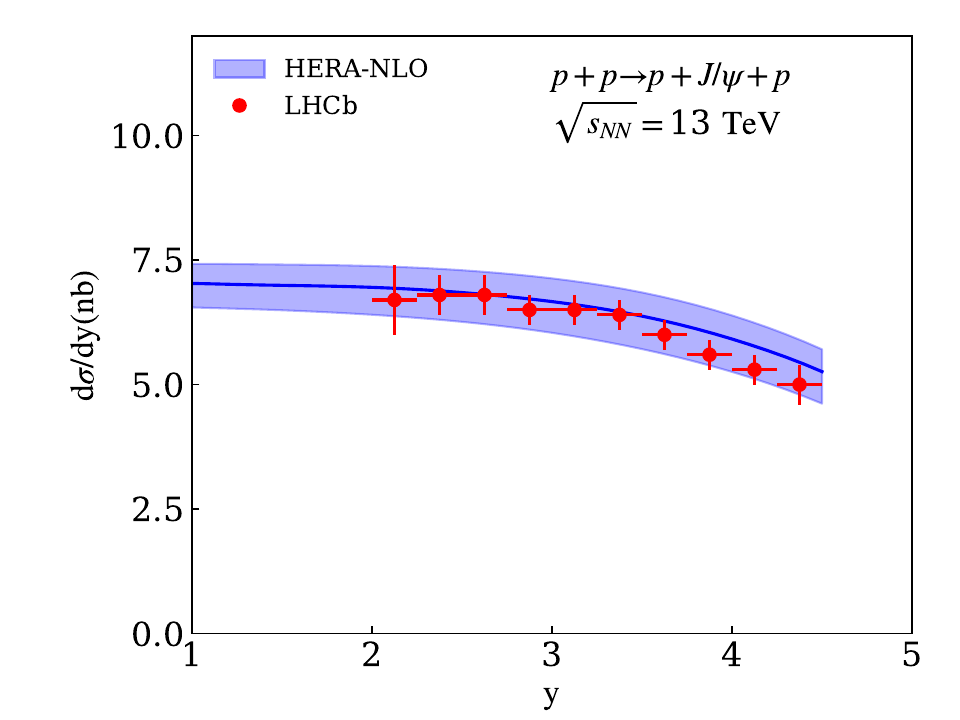}
	\caption{Exclusive production of $J/\psi$  as a function of rapidity in proton-proton collisions at LHCb and comparison with experimental data from LHCb\cite{LHCb:2014acg, LHCb:2016oce}. The gluon density is HEAR15 set \cite{H1:2015ubc}.}
\end{figure}

Furthermore, the prediction of exclusive $J/\psi$ in proton-proton collisions at NICA $\sqrt{s_{NN}}$ = 24 GeV and RHIC $\sqrt{s_{NN}}$ = 200 GeV are
 also performed using the same sets of gluon density. Our
results are presented in Fig.~4. At RHIC and NICA energies the survival factors are set to unity because the collision energy is much lower than for LHCb in
 accordance with \cite{Jones:2016icr} results. The gluon density uncertainties are small at large x region and  the GPDs generate small uncertainties
  of $J/\psi$ production cross section at RHIC and NICA energies.
This means that the exclusive vector meson production in proton-proton collisions can be employed to obtain an additional
constrain for the uncertainties of gluon density.
In addition, the exclusive $J/\psi$ cross section at proton-proton collisions can help to simulate the detector system of NICA.
One can apply our predictions to estimate the event number accepted in the detectors.

Finally, we can conclude that the GPD approach can be employed to calculate the exclusive $J/\psi$ production at proton-proton collisions
at LHCb and lower RHIC and NICA energies. The survival factors and the equivalent photon approximation are important in predicting the heavy vector meson production of these reactions.

\begin{figure}[h!]
	\includegraphics[width=6.9cm]{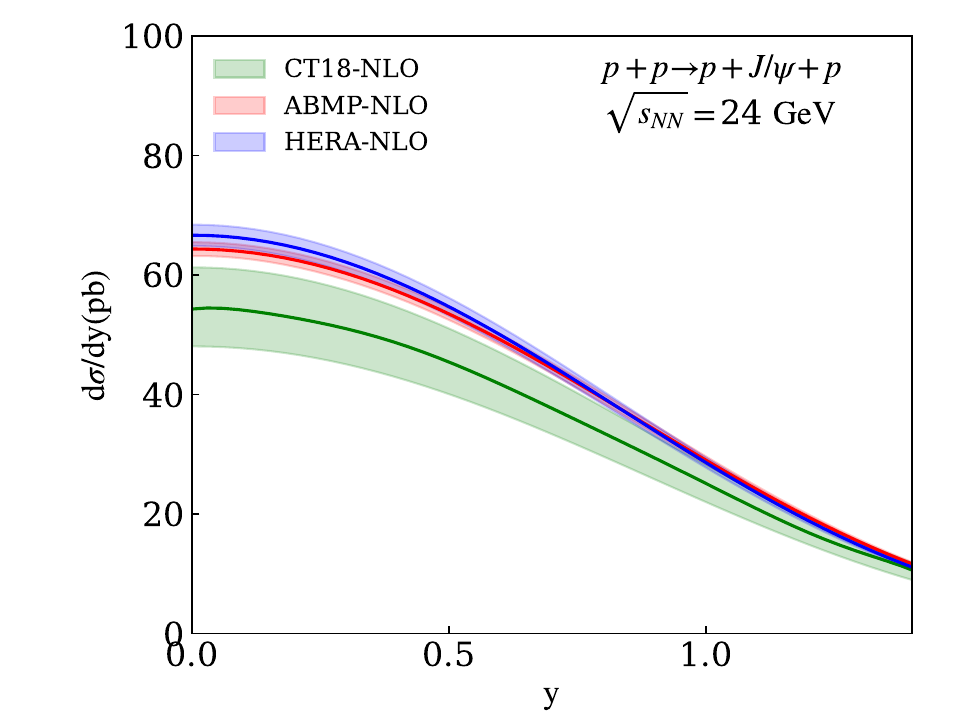}
	\includegraphics[width=6.9cm]{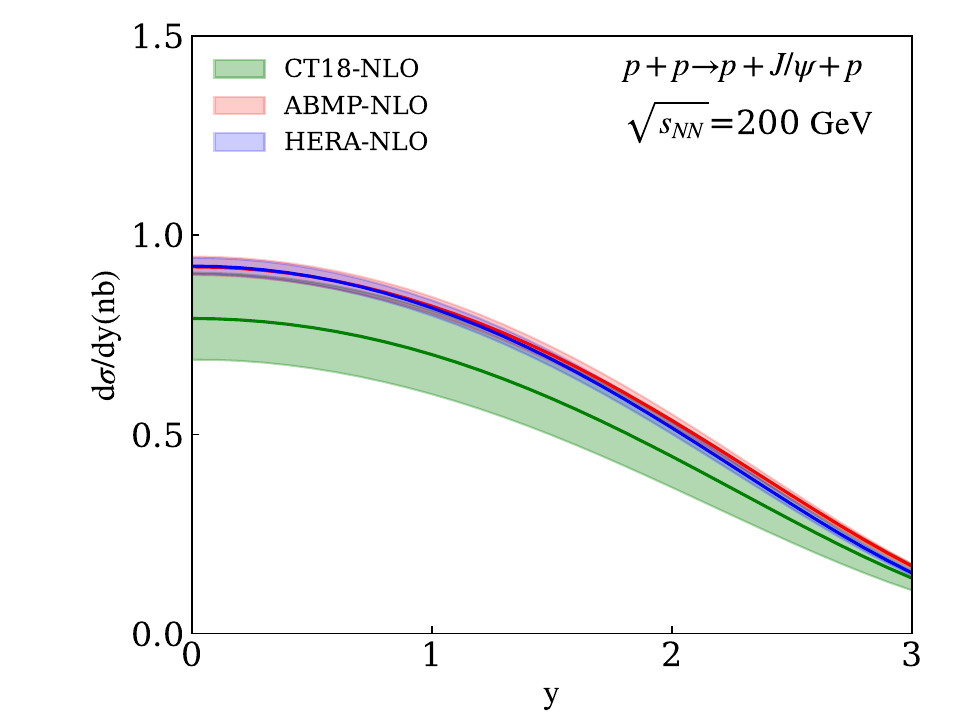}
	\caption{Exclusive production of $J/\psi$ as a function of rapidity at proton-proton collisions at NICA and RHIC energy regions.}
\end{figure}

\section{Summary}
In this paper, we study the exclusive $J/\psi$ production at proton-proton collisions within GPD approach.  The equivalent photon approximation is adopted 
to estimate the photon flux of proton and the survival factors
are also employed in the calculation. Three sets of gluon density are implemented in calculations of the gluon GPDs  within GK model. Our predictions 
give a good description to the LHCb experimental data on the  exclusive $J/\psi$ rapidity distributions. 
 The uncertainties of $J/\psi$
 cross section are also given.  It can be seen that the gluon density uncertainties can affect the rapidity distribution. At the same time, the $J/\psi$ rapidity distributions can be adopted to constrain  the gluon density and corresponding GPDs uncertainties.  The survival factors are essential ingredient  in the calculations of $J/\psi$ production in the proton-proton ultraperipheral collisions. We use in our calculations results from  \cite{Jones:2016icr}.
It it is important to study properties of the survival factors at high energy more carefully in the future.

The exclusive $J/\psi$ production at RHIC and NICA are also investigated and the prediction of $J/\psi$ cross section as a function of rapidity are found for the same
 sets of gluon density. The survival factors at RHIC and NICA is equal to unity in our calculation since the collision energy is lower than LHCb in accordance with \cite{Jones:2016icr}.  It can be seen that the
  uncertainties of $J/\psi$ production cross section at RHIC and NICA are smaller than LHCb in our prediction since the gluon density uncertainties are much smaller at low energy. We conclude that
the $J/\psi$ production at RHIC and NICA can be employed to study the properties of gluon GPD at large $x$ region.
\section*{Acknowledgment}
Y. P. Xie gives many thanks to Victor Goncalves and Ya-Jin Zhou for discussion.
This work is partially supported by  National Key R\&D Program of China (Grant No. 2024YFE0109800),
 the NFSC grant (Grant No. 12293061) and CAS president's international fellowship initiative (Grant No. 2025VMA0005).


\begin{thebibliography}{}
\bibitem{upc}
C. A. Bertulani and G. Baur, { Phys. Rep.} {\bf 163}, 299 (1988); F.~Krauss, M.~Greiner and G.~Soff,
Prog.\ Part.\ Nucl.\ Phys.\  {\bf 39}, 503 (1997);
\bibitem{upc02}
C.~A. Bertulani, S.~R.~Klein and J.~Nystrand, Ann. Rev. Nucl. Part. Sci. {\bf 55},
271 (2005);


\bibitem{twophotons}
G.~Baur, K.~Hencken, D.~Trautmann, S.~Sadovsky and Y.~Kharlov,
Phys. Rept. \textbf{364}, 359-450 (2002)
[arXiv:hep-ph/0112211 [hep-ph]].


\bibitem{ALICE:2013wjo}
E.~Abbas \textit{et al.} [ALICE],
Eur. Phys. J. C \textbf{73}, no.11, 2617 (2013)
[arXiv:1305.1467 [nucl-ex]].

J.~Adam \textit{et al.} [ALICE],
JHEP \textbf{09}, 095 (2015)
[arXiv:1503.09177 [nucl-ex]].

S.~Acharya \textit{et al.} [ALICE],
Eur. Phys. J. C \textbf{79}, no.5, 402 (2019)
[arXiv:1809.03235 [nucl-ex]].

S.~Acharya \textit{et al.} [ALICE],
Phys. Rev. D \textbf{108}, no.11, 112004 (2023)
[arXiv:2304.12403 [nucl-ex]].


\bibitem{LHCb:2013nqs}
R.~Aaij \textit{et al.} [LHCb],
J. Phys. G \textbf{40}, 045001 (2013)
[arXiv:1301.7084 [hep-ex]].


\bibitem{LHCb:2014acg}
R.~Aaij \textit{et al.} [LHCb],
J. Phys. G \textbf{41}, 055002 (2014)
[arXiv:1401.3288 [hep-ex]].

\bibitem{LHCb:2016oce}
[LHCb],
LHCb-CONF-2016-007.

\bibitem{PHENIX:2009xtn}
S.~Afanasiev \textit{et al.} [PHENIX],
Phys. Lett. B \textbf{679}, 321-329 (2009)
[arXiv:0903.2041 [nucl-ex]].


\bibitem{LHCUPC}
 A.~J.~Baltz {\it et al.},
Phys.\ Rept.\  {\bf 458}, 1 (2008);
 J.~G.~Contreras and J.~D.~Tapia Takaki,
Int.\ J.\ Mod.\ Phys.\ A {\bf 30}, 1542012 (2015).


\bibitem{GPD-review01}
K.~Goeke, M.~V.~Polyakov and M.~Vanderhaeghen,
Prog. Part. Nucl. Phys. \textbf{47}, 401-515 (2001)
[arXiv:hep-ph/0106012 [hep-ph]].




\bibitem{GPD-review02}
M.~Diehl,
Phys. Rept. \textbf{388}, 41-277 (2003)
[arXiv:hep-ph/0307382 [hep-ph]];

 A. V. Belitsky and A. V. Radyushkin, Phys. Rept. 418 (2005) 1 [hep-ph/0504030]


\bibitem{Muller:1994ses}
D.~M\"uller, D.~Robaschik, B.~Geyer, F.~M.~Dittes and J.~Ho\v{r}ej\v{s}i,
Fortsch. Phys. \textbf{42}, 101-141 (1994).

\bibitem{Ji:1996ek}
X.~D.~Ji,
Phys. Rev. Lett. \textbf{78}, 610-613 (1997);
Phys. Rev. D \textbf{55}, 7114-7125 (1997)
[arXiv:hep-ph/9609381 [hep-ph]].


\bibitem{Radyushkin:1997ki}
A.~V.~Radyushkin,
Phys. Lett. B \textbf{385}, 333-342 (1996)
[arXiv:hep-ph/9605431 [hep-ph]].
A.~V.~Radyushkin,
Phys. Rev. D \textbf{56}, 5524-5557 (1997).

\bibitem{dvcs}
J. J.~C.~Collins, L.~Frankfurt and M.~Strikman,
Phys. Rev. D \textbf{56}, 2982-3006 (1997)
[arXiv:hep-ph/9611433 [hep-ph]].

\bibitem{TCS}
E.~R.~Berger, M.~Diehl and B.~Pire,
Eur. Phys. J. C \textbf{23}, 675-689 (2002);
D.~Mueller, B.~Pire, L.~Szymanowski and J.~Wagner,
Phys. Rev. D \textbf{86}, 031502 (2012);
M.~Bo\"er, M.~Guidal and M.~Vanderhaeghen,
Eur. Phys. J. A \textbf{51}, no.8, 103 (2015).

\bibitem{Xie:2022vvl}
Y.~P.~Xie and V.~P.~Goncalves,
Phys. Lett. B \textbf{839}, 137762 (2023)
[arXiv:2212.07657 [hep-ph]].
Eur. Phys. J. C \textbf{83}, no.6, 528 (2023);
Eur. Phys. J. C \textbf{84}, no.9, 880 (2024)
[arXiv:2409.11213 [hep-ph]].

\bibitem{dvmp1}
M.~Vanderhaeghen, P.~A.~M.~Guichon and M.~Guidal,
Phys. Rev. D \textbf{60}, 094017 (1999)
[arXiv:hep-ph/9905372 [hep-ph]].
\bibitem{Goloskokov:2022mdn}
S.~V.~Goloskokov, Y.~P.~Xie and X.~Chen,
Chin.Phys.C 46 (2022) 12, 123101
[arXiv:2206.06547 [hep-ph]].
S.~V.~Goloskokov, Y.~P.~Xie and X.~Chen,
Commun. Theor. Phys. \textbf{75}, no.6, 065201 (2023)
[arXiv:2209.14493 [hep-ph]].

\bibitem{Pire:2017tvv}
B.~Pire, L.~Szymanowski and J.~Wagner,
Phys. Rev. D \textbf{95}, no.11, 114029 (2017)
[arXiv:1705.11088 [hep-ph]].
B.~Pire and L.~Szymanowski,
Phys. Rev. D \textbf{96}, no.11, 114008 (2017)
[arXiv:1711.04608 [hep-ph]].


\bibitem{PRD68}
N.N. Nikolaev and B.G. Zakharov, Z. Phys. C 49 (1991) 607.
A.H. Mueller, Nucl. Phys. B 415, 373 (1994).
H.~Kowalski and D.~Teaney,
Phys. Rev. D \textbf{68}, 114005 (2003)
[arXiv:hep-ph/0304189 [hep-ph]].
\bibitem{PRD74}
H.~Kowalski, L.~Motyka and G.~Watt,
Phys. Rev. D \textbf{74}, 074016 (2006)
[arXiv:hep-ph/0606272 [hep-ph]].
A.~H.~Rezaeian, M.~Siddikov, M.~Van de Klundert and R.~Venugopalan,
Phys. Rev. D \textbf{87}, no.3, 034002 (2013)
[arXiv:1212.2974 [hep-ph]].

\bibitem{PRD78}
G.~Watt and H.~Kowalski,
Phys. Rev. D \textbf{78}, 014016 (2008)
[arXiv:0712.2670 [hep-ph]].
A.~H.~Rezaeian and I.~Schmidt,
Phys. Rev. D \textbf{88}, 074016 (2013)
[arXiv:1307.0825 [hep-ph]].

\bibitem{Lappi:2013am}
T.~Lappi and H.~Mantysaari,
Phys. Rev. C \textbf{87}, no.3, 032201 (2013)
[arXiv:1301.4095 [hep-ph]].

\bibitem{Goncalves:2014swa}
V.~P.~Gon\c{c}alves, B.~D.~Moreira and F.~S.~Navarra,
Phys. Lett. B \textbf{742}, 172-177 (2015)
[arXiv:1408.1344 [hep-ph]].


\bibitem{Xie:2016ino}
Y.~P.~Xie and X.~Chen,
Eur. Phys. J. C \textbf{76}, no.6, 316 (2016)
[arXiv:1602.00937 [hep-ph]].

\bibitem{Xie:2018tng}
Y.~P.~Xie and X.~Chen,
Int. J. Mod. Phys. A \textbf{33}, no.14n15, 1850086 (2018)
[arXiv:1805.11480 [hep-ph]].


\bibitem{Goloskokov:2024egn}
S.~V.~Goloskokov, Y.~P.~Xie and X.~Chen,
Phys. Rev. D \textbf{110}, no.7, 076029 (2024)
[arXiv:2408.05800 [hep-ph]].

\bibitem{Khoze:2013jsa}
V.~A.~Khoze, A.~D.~Martin and M.~G.~Ryskin,
Eur. Phys. J. C \textbf{74}, no.2, 2756 (2014)
[arXiv:1312.3851 [hep-ph]].

\bibitem{Jones:2013pga}
S.~P.~Jones, A.~D.~Martin, M.~G.~Ryskin and T.~Teubner,
JHEP \textbf{11}, 085 (2013)
[arXiv:1307.7099 [hep-ph]].

\bibitem{Jones:2016icr}
S.~P.~Jones, A.~D.~Martin, M.~G.~Ryskin and T.~Teubner,
J. Phys. G \textbf{44}, no.3, 03LT01 (2017)
[arXiv:1611.03711 [hep-ph]].



\bibitem{Xing:2020hwh}
H.~Xing, C.~Zhang, J.~Zhou and Y.~J.~Zhou,
JHEP \textbf{10}, 064 (2020)
[arXiv:2006.06206 [hep-ph]].

\bibitem{Zha:2020cst}
W.~Zha, J.~D.~Brandenburg, L.~Ruan, Z.~Tang and Z.~Xu,
Phys. Rev. D \textbf{103}, no.3, 033007 (2021)
[arXiv:2006.12099 [hep-ph]].

\bibitem{yajin-zhou}
Private communication with Ya-Jin Zhou.



\bibitem{Sorin:2011zz}
A.~Sorin, V.~Kekelidze, A.~Kovalenko, R.~Lednicky, I.~Meshkov and G.~Trubnikov,
Nucl. Phys. A \textbf{855}, 510-513 (2011).

\bibitem{STAR:2004jhn}
H.~Caines \textit{et al.} [STAR],
J. Phys. G \textbf{30}, S61-S73 (2004).

\bibitem{PHENIX:2004vcz}
K.~Adcox \textit{et al.} [PHENIX],
Nucl. Phys. A \textbf{757}, 184-283 (2005)
[arXiv:nucl-ex/0410003 [nucl-ex]].

\bibitem{gk06}
S.~V.~Goloskokov and P.~Kroll,
Eur. Phys. J. C \textbf{42}, 281-301 (2005)
[arXiv:hep-ph/0501242 [hep-ph]].
Eur. Phys. J. C \textbf{50}, 829-842 (2007)
[arXiv:hep-ph/0611290 [hep-ph]].
\bibitem{gk08}
S.~V.~Goloskokov and P.~Kroll,
Eur. Phys. J. C \textbf{59}, 809-819 (2009)
[arXiv:0809.4126 [hep-ph]].

\bibitem{gk09}
S.~V.~Goloskokov and P.~Kroll,
Eur. Phys. J. C \textbf{65}, 137-151 (2010)
[arXiv:0906.0460 [hep-ph]].
S.~V.~Goloskokov and P.~Kroll,
Eur. Phys. J. A \textbf{47}, 112 (2011)
[arXiv:1106.4897 [hep-ph]].





\bibitem{gk07}
S.~V.~Goloskokov and P.~Kroll,
Eur. Phys. J. C \textbf{53}, 367-384 (2008)
[arXiv:0708.3569 [hep-ph]].


\bibitem{mus99}
I.~V.~Musatov and A.~V.~Radyushkin,
Phys. Rev. D \textbf{61}, 074027 (2000)
[arXiv:hep-ph/9905376 [hep-ph]].


\bibitem{Flett:2021ghh}
D.~Y.~Ivanov, L.~Szymanowski and G.~Krasnikov,
JETP Lett. \textbf{80}, 226-230 (2004)
[erratum: JETP Lett. \textbf{101}, no.12, 844 (2015)]
[arXiv:hep-ph/0407207 [hep-ph]].
C.~A.~Flett, J.~A.~Gracey, S.~P.~Jones and T.~Teubner,
JHEP \textbf{08}, 150 (2021)
[arXiv:2105.07657 [hep-ph]].
C.~A.~Flett, J.~P.~Lansberg, S.~Nabeebaccus, M.~Nefedov, P.~Sznajder and J.~Wagner,
Phys. Lett. B \textbf{859}, 139117 (2024)
doi:10.1016/j.physletb.2024.139117
[arXiv:2409.05738 [hep-ph]].



\bibitem{Hou:2019qau}
T.~J.~Hou, K.~Xie, J.~Gao, \textit{et al.}
[arXiv:1908.11394 [hep-ph]].

\bibitem{Alekhin:2018pai}
S.~Alekhin, J.~Bl\"umlein and S.~Moch,
Eur. Phys. J. C \textbf{78}, no.6, 477 (2018)
[arXiv:1803.07537 [hep-ph]].

\bibitem{H1:2015ubc}
H.~Abramowicz \textit{et al.} [H1 and ZEUS],
Eur. Phys. J. C \textbf{75}, no.12, 580 (2015)
[arXiv:1506.06042 [hep-ex]].




\end{thebibliography}
\end{document}